\documentclass[twocolumn,superscriptaddress,amsmath,amssymb,showpacs,prl]{revtex4-1}

\usepackage{graphicx}

\usepackage{color}
\renewcommand{\phi}{\varphi}

\begin{document}

\title{Two-step nucleation of the Earth's inner core}

\author{Yang Sun}
    \email{ys3339@columbia.edu}
	\affiliation{Department of Applied Physics and Applied Mathematics, Columbia University, New York, NY 10027, USA}
\author{Feng Zhang}
	\affiliation{Ames Laboratory, US Department of Energy, Ames, Iowa 50011, USA}
\author{Mikhail I. Mendelev}
    \email{mikhail.mendelev@gmail.com}
	\affiliation{Ames Laboratory, US Department of Energy, Ames, Iowa 50011, USA}
\author{Renata M. Wentzcovitch}
	\affiliation{Department of Applied Physics and Applied Mathematics, Columbia University, New York, NY 10027, USA}
	\affiliation{Department of Earth and Environmental Sciences, Columbia University, New York, NY 10027, USA}
	\affiliation{Lamont–Doherty Earth Observatory, Columbia University, Palisades, NY 10964, USA}
\author{Kai-Ming Ho}
	\affiliation{Department of Physics, Iowa State University, Ames, Iowa 50011, USA}

\date{Jan. 8, 2022}

\begin{abstract}

\textbf{The Earth's inner core started forming when molten iron cooled below the melting point. However, the nucleation mechanism, which is a necessary step of crystallization, has not been well understood. Recent studies found it requires an unrealistic degree of undercooling to nucleate the stable hexagonal close-packed (hcp) phase of iron that is unlikely to be reached under core conditions and age. This contradiction is referred as the inner core nucleation paradox. Using a persistent-embryo method and molecular dynamics simulations, we demonstrate that the metastable body-centered cubic (bcc) phase of iron has a much higher nucleation rate than the hcp phase under inner-core conditions. Thus, the bcc nucleation is likely to be the first step of inner core formation instead of direct nucleation of the hcp phase. This mechanism reduces the required undercooling of iron nucleation, which provides a key factor to solve the inner-core nucleation paradox. The two-step nucleation scenario of the inner core also opens a new avenue for understanding the structure and anisotropy of the present inner core.}

\end{abstract}

\maketitle

The core plays a key role in the Earth’s evolution. The present core contains two major parts, a solid inner core and a liquid outer core. Iron dominates both parts with a small amount of light elements \cite{1}. The solid core is generally believed to be hcp iron, while the possible existence of bcc iron has also been suggested \cite{2,3,4,5}. The growth of the solid inner core is believed to be the major driving force of the present geodynamo, providing the main power source for convection in the liquid core \cite{6, 7}. Despite its importance, the initial formation of the solid core, which directly relates to its thermal evolution and Earth’s history, is far from being completely understood \cite{8,9,10,11,12}. Most of Earth’s thermal history models assume that the inner core started to crystallize when molten iron cooled right below its melting temperature at the Earth’s center \cite{7}. However, in practice, nucleation does not happen at the melting point but requires some undercooling because of the formation of a solid-liquid interface (SLI) that accompanies it. While the bulk solid phase is thermodynamically favored, the SLI costs energy. These two factors lead to a nucleation barrier $\Delta$G, which is described in classical nucleation theory (CNT) \cite{13} as 
\begin{equation}
\Delta G = N \Delta \mu + A \gamma ,
\end{equation}
where $N$ is the nucleus size, $\Delta \mu\ (<0)$ is the free energy difference between the bulk solid and liquid, $\gamma\ (>0)$ is the SLI free energy, and $A$ is the SLI area. The liquid must be cooled sufficiently below the melting temperature to overcome the free energy barrier during thermal fluctuations. After considering this mechanism, it was found that a very large undercooling of $\sim$1000 K is required for the nucleation of hcp iron in the Earth's core \cite{14}. However, considering the slow cooling rate of $\sim$100 K/Gyr throughout the core history \cite{15}, it is impossible to reach such a large degree of undercooling inside the Earth within the inner core’s age. This ``inner core nucleation paradox", recently described by Huguet \textit{et al.} \cite{14}, strongly challenges the current understanding of the inner core formation process. While Huguet \textit{et al.}'s argument relies on a few estimations of thermodynamic quantities, Davies \textit{et al.} also confirmed the paradox with atomic-scale simulations \cite{16}. Even considering the effect of light elements on the nucleation process, it still requires 675 K undercooling to nucleate hcp iron, nearly impossible to reach in the Earth core \cite{16}.

CNT was proposed more than a century ago, and its formalism is the most widely used to describe nucleation phenomena nowadays. The simplest scenario in CNT assumes a single nucleation pathway where only the nucleus of the thermodynamically stable phase forms and grows towards the bulk phase. This was the situation considered in \cite{14, 15}, where the authors assumed that the melt in the Earth’s core crystallized directly into the hcp phase. Recent studies have shown that nucleation can be a multi-step process that includes multiple intermediate stages and phases \cite{17,18,19}. While the CNT concept of nucleus formation is still valid under these situations, phase competition must be considered \cite{18, 19}. Therefore, instead of the scenario described above, we can consider a complex process where nucleation is facilitated by forming an intermediate phase with a high nucleation rate. For example, it has been observed that the bcc phase can nucleate before the face-centered cubic (fcc) or hcp phases in a few alloys where the fcc/hcp phase is the most stable one \cite{20,21,22,23,24}. Could the bcc phase also facilitate hcp iron nucleation and relate to the inner core nucleation paradox? Making a quantitative prediction on such complex nucleation processes is a challenging problem. In addition to the extreme conditions in the core, nucleation involves microscopic length scales that are extremely hard to probe in real-time, even with state-of-the-art measurements \cite{25}. Hence, it requires computer simulations, particularly large-scale molecular dynamics (MD), to reproduce the temporal evolution of the liquid into the crystal \cite{26}. Unfortunately, nucleation under Earth’s core conditions is a rare event that occurs on the geological time scale, far beyond the reach of conventional MD simulations. Besides, large-scale MD simulations require semi-empirical potentials to describe atomic interactions, and the outcome may depend heavily on the potential’s quality \cite{27}. In this work, we assess the inner-core nucleation process with the account of competition between bcc and hcp phases during the nucleation process using the persistent-embryo method (PEM) \cite{28} to overcome the significant time limitation in conventional MD simulation of nucleation. 

\begin{figure}
\includegraphics[width=0.8\textwidth]{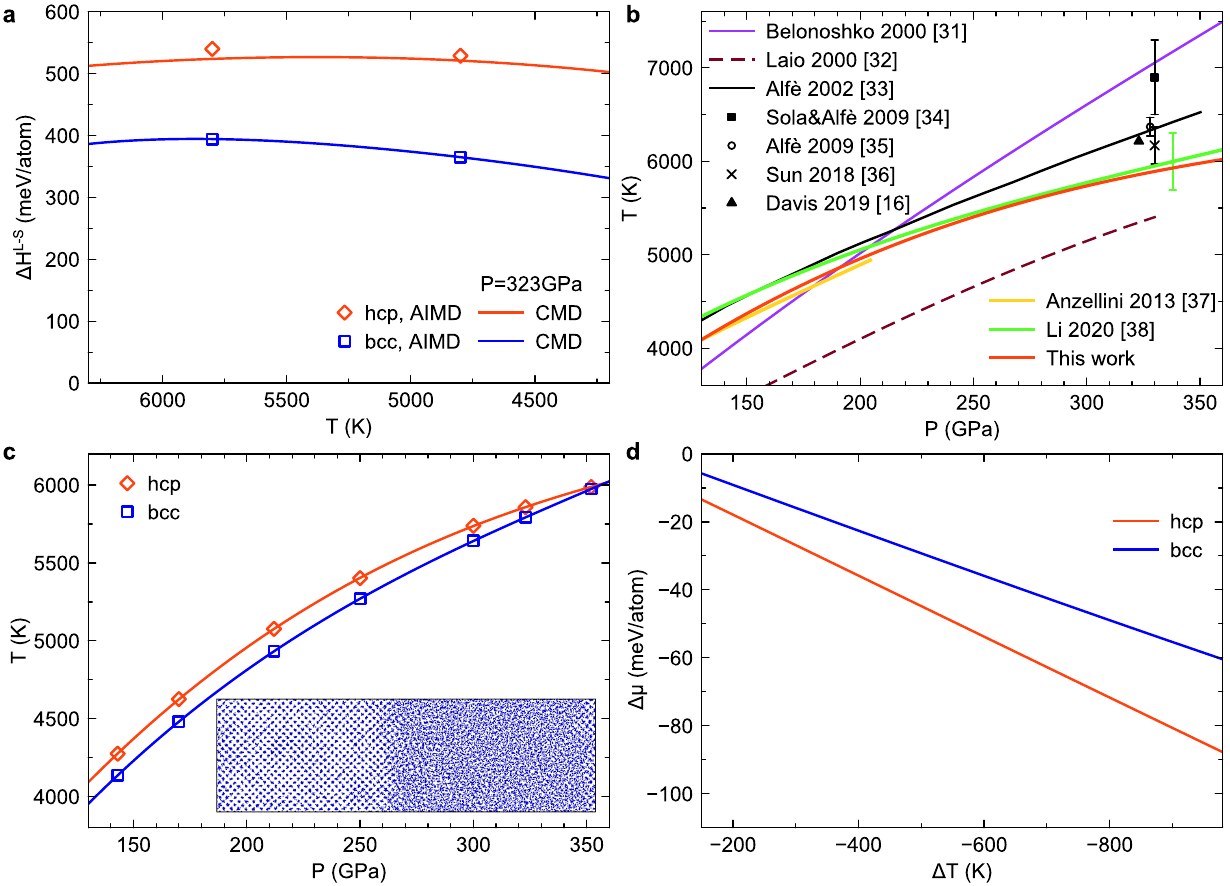}
\caption{\label{fig:fig1} \textbf{Melting curves and thermodynamic properties of hcp and bcc iron.} a, The latent heats of hcp and bcc iron at 323 GPa from AIMD and CMD with the developed semi-empirical potential. b, Comparison between the current melting curve of hcp iron and several others obtained by previous simulations and experiments\cite{16, 31,32,33,34,35,36,37,38}. c, Melting curves of hcp and bcc iron from CMD with the developed semi-empirical potential. The inset shows a projected snapshot of the bcc solid-liquid coexistence simulation. d, Change in bulk free energy upon solidification (nucleation driving force) at 323 GPa.}
\end{figure}

\textbf{Melting curve of hcp and bcc phases.}
The nucleation rate of hcp iron was previously estimated \cite{14} based on the driving force and the SLI free energy obtained in \cite{29} with the semi-empirical potential developed by Ackland \textit{et al.} \cite{30}. However, this potential was developed to simulate iron at ambient conditions such that no high-pressure data were used in the potential development \cite{30}. In the present study, we developed a potential explicitly considering its application at Earth's core conditions. The details of the potential development are in Supplementary Information. One of the vital target properties in the potential development is the latent heat $\Delta H^{L-S}$, because along with the melting temperature it defines the driving force for the solidification. Figure \ref{fig:fig1}a shows excellent agreement between the latent heat calculated using the developed potential and \textit{ab initio} MD (AIMD) for both the hcp and bcc phases. Besides, elastic properties and liquid structure predicted with the developed potential also agree well with those calculated with AIMD (see Supplementary Information), making this potential suitable for simulations of the iron crystallization process under inner-core conditions. 

\onecolumngrid

\clearpage

\twocolumngrid

In Fig. \ref{fig:fig1}b, we compare the current melting curve of the hcp phase with previous measurements and simulation results \cite{16,31,32,33,34,35,36,37,38}. This is especially important because melting curves from previous classical MD (CMD) simulations deviate considerably from each other \cite{31, 32}, which points out the importance of employing a thoroughly developed semi-empirical potential. The melting temperatures in the present study were determined at several pressures using the solid-liquid coexistence approach \cite{39}. The current melting curve agrees well with the experimental curve obtained using fast x-ray diffraction in the laser-heated diamond anvil cell in the pressure range between 130 GPa and 200 GPa \cite{37}. It also reasonably agrees with the recent estimation of melting boundary from shock compression measurements in the higher pressure range from 250 GPa to 360 GPa \cite{38}.Compared to previous simulations, our current melting curve provides the closest agreement to the recent high-pressure experiments for the hcp phases. It also validates the calculations of latent heat, which is directly related to the slope of the melting curve according to the Clausius-Clapeyron equation.

The melting curves of hcp and bcc phases are compared in Figure 1c. The hcp phase has higher melting temperatures in the pressure range from 130 GPa to 330 GPa. Thus, the hcp phase is thermodynamically stable, while the bcc phase is metastable in the range from the core-mantle boundary to the inner core boundary. Interestingly, the melting points of hcp and bcc phases are predicted to be very close after 330 GPa and crossover at ~360 GPa, which suggests a similar free energy of bcc and hcp phases at pressures near the inner core center. 

In the present study, we chose to conduct simulations of nucleation at 323 GPa, the pressure at the inner core boundary. We used the Gibbs-Helmholtz equation and MD simulation to calculate the free energy difference $\Delta \mu$ between the bulk solid and liquid as described in \cite{40}. Figure \ref{fig:fig1}d shows $\Delta \mu$ as a function of undercooling with respect to the hcp melting temperature, $\Delta T = T-T_m^{hcp}$, for both the hcp and bcc phases. The absolute value of $\Delta \mu$ for the hcp phase is always larger than that for the bcc phase. 

\begin{figure}
\includegraphics[width=0.45\textwidth]{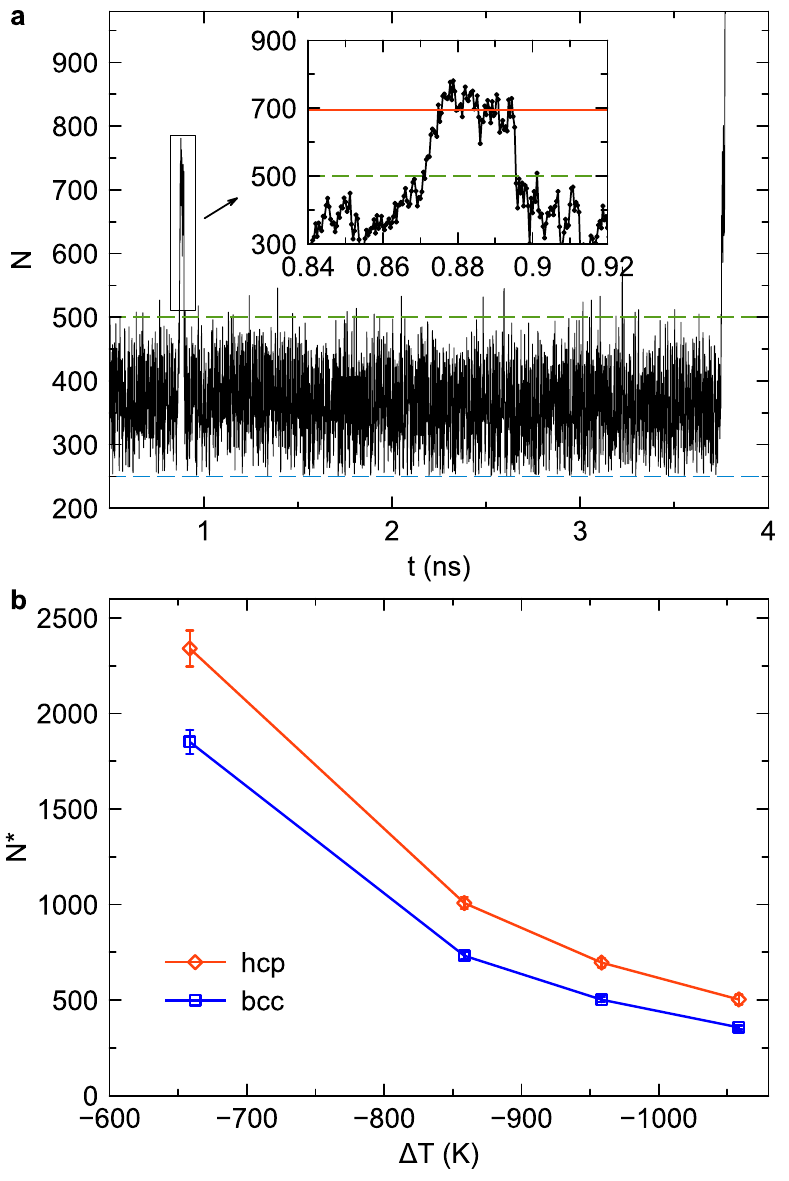}
\caption{\label{fig:fig2} \textbf{PEM-MD simulation and critical nucleus size.} a, The nucleus size versus time of a PEM-MD trajectory at $\Delta$T=-958 K (i.e., T=4900 K). The blue dashed line shows the size of persistent embryo, $N_0$, and the green dashed line indicates the threshold for aping removal, $N_{sc}$ (see Method). The inset enlarges the plateaus at the critical size. The red line indicates the plateaus to determine the critical nucleus size, $N^*$. b, The critical nucleus size as a function of undercooling temperature for hcp and bcc phases. }
\end{figure}

\textbf{Nucleation of hcp and bcc phases}
According to the CNT \cite{13}, the nucleation barrier $\Delta G^*$ is the key quantity to determine the nucleation rate. It can be computed \cite{28} as 
\begin{equation}
\Delta G^* = \frac{1}{2} \lvert \Delta  \mu \rvert N^* ,
\end{equation}
where $N^\ast$ is the critical nucleus size. To obtain $N^\ast$, we employ the PEM \cite{28}, which invokes the central CNT concept that homogeneous nucleation happens via the formation of a critical nucleus in the undercooled liquid. Figure 2a shows a typical result of the PEM-MD simulation. The plateau on the MD trajectory indicates the appearance of the critical nucleus (see technical details in Methods). The critical nucleus sizes of both the hcp and bcc phases at several moderate undercooling temperatures are shown in Fig. 2b. The hcp phase shows a systematically larger critical nucleus size than the bcc phase. Figure 3a shows the free energy barriers $\Delta G^*$ of both bcc and hcp phases computed using Eq. (2). The hcp phase has a larger nucleation barrier than the bcc phase at all undercooling temperatures considered here, although the hcp phase has a larger bulk driving force $\Delta \mu$ in Fig. 1d. To explain this, we compare the SLI free energies of these phases obtained from the PEM simulation. Figure 3b shows that this quantity is much larger for the hcp phase. This is consistent with a previous study which suggests that bcc metals show lower interface free energy than hcp metals at ambient conditions \cite{41}. Because the $\Delta G^*$ scales with $\gamma^3$, the difference in $\gamma$ can significantly change the ratio of the nucleation barriers. The temperature dependences of the $\gamma$ obtained for both hcp and bcc phases are almost linear as shown in Fig. 3b, which is similar to the ones found for Ni and Al in \cite{42}. Therefore, the SLI free energy can be linearly extrapolated to smaller undercooling (higher temperatures), where the critical nucleus size is too large to be simulated directly. In Fig. 3b we further compare the $\gamma$ of hcp with previous data. Davies \textit{et al.} computed the $\gamma$ based on a different semi-empirical potential and the seeding technique \cite{16}. While the temperature dependence was not considered in that work, the value of $\gamma$ is very similar to our data. In contrast, the $\gamma$ determined in Ref.\cite{29} highly deviates from our data and Ref. \cite{16}. This can be attributed to the fact that the empirical potential used in Ref.\cite{29} was not designed to simulate iron under the Earth’s core conditions.

\begin{figure}
\includegraphics[width=0.45\textwidth]{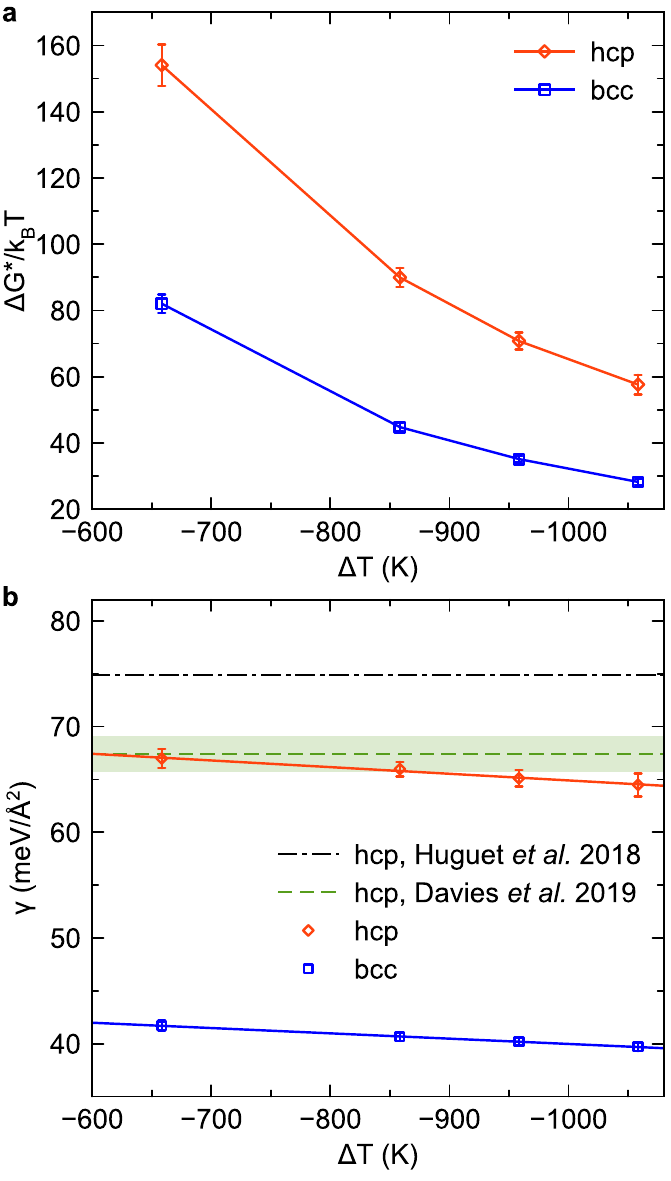}
\caption{\label{fig:fig3} \textbf{Temperature dependence of the free energy barrier and SLI free energy.} a, Free energy barrier $\Delta G^*$ as a function of undercooling temperature for hcp and bcc at 323 GPa. b, SLI free energy of hcp and bcc phases at 323 GPa. The dashed dot line is from Ref.\cite{14}. The dashed line with the confidence interval (green band) is from Ref.\cite{16}.}
\end{figure}

The nucleation rate, $J$, can be calculated as $J = \kappa \ \text{exp}(-\Delta G^*/ k_B T)$, where $k_B$ is the Boltzmann constant, and $\kappa$ is a kinetic prefactor. As demonstrated by Auer and Frenkel \cite{44}, the kinetic prefactor can be determined from MD simulations based on the fluctuations of the nucleus size around the critical value (see Supplementary Information). Thus we obtained all essential parameters to compare the bcc and hcp nucleation rates with PEM and MD. Figure 4a shows that the bcc phase has a much higher nucleation rate than the hcp phase in a broad undercooling regime. For example, at an undercooling of 660 K, the nucleation rate of the bcc phase is \textit{31 orders of magnitudes} higher than that of the hcp phase. With such a vast difference, the bcc phase should always nucleate much quicker than the hcp phase under the Earth’s core conditions.

\begin{figure}
\includegraphics[width=0.45\textwidth]{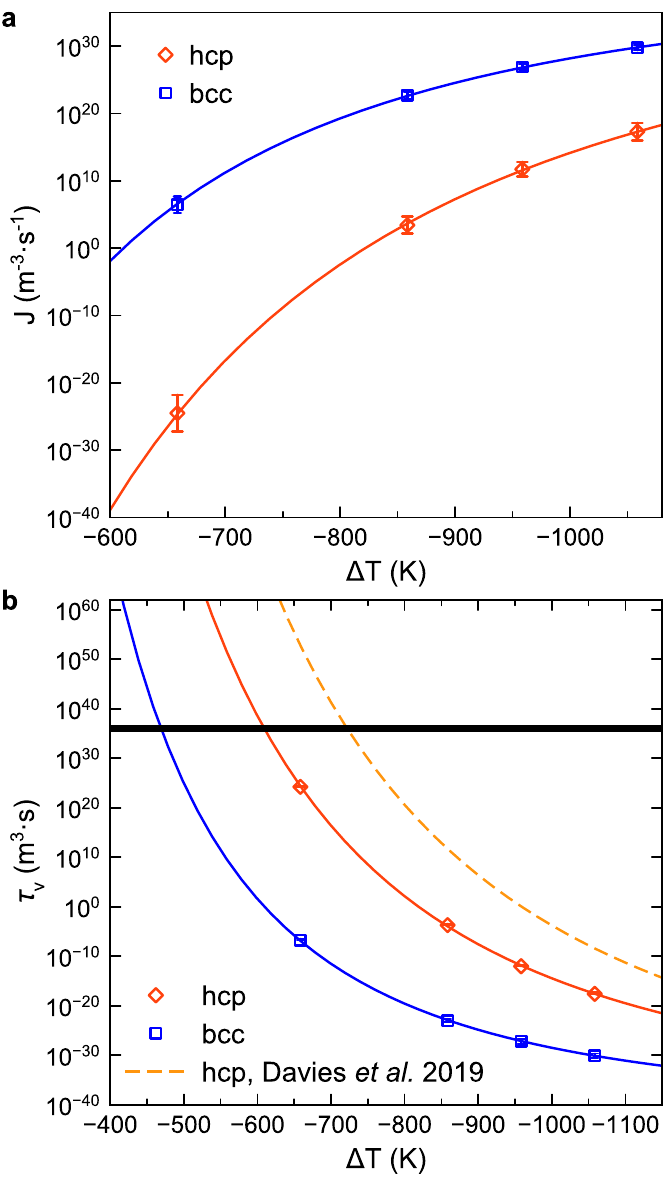}
\caption{\label{fig:fig4} \textbf{Nucleation rate and waiting time at 323 GPa.}  a, Nucleation rate as a function of undercooling for the hcp and bcc phases. b, Waiting time as a function of undercooling. The dashed line is from Ref. \cite{16}. The thick black line indicates the range of nucleation waiting time $\tau_v$ in the Earth’s core.}
\end{figure}

Using the obtained nucleation rate, we are now able to estimate the nucleation waiting time. Because the critical nucleus only has half chance to grow at the top of the nucleation barrier, the waiting time in a fixed volume can be expressed as $\tau_v=\frac{1}{2J}$ \cite{16}. As shown in Fig. \ref{fig:fig4}b, the present waiting time of hcp is smaller than the previous estimation by Davies \textit{et al.} \cite{16}. This is mainly because the semi-empirical potential used in \cite{16} shows a different melting point and bulk free energy difference than the present values (see Supplementary Information). The reachable $\tau_v$ in the Earth’s core can be estimated as follows. The nucleation incubation time is approximated as one billion years, probably the upper limit of plausible inner core age \cite{16}. The volumes of the inner core and the entire core are $7.6\times{10}^{18}$ $m^3$ to $1.8\times{10}^{20}$ $m^3$. Therefore, $\tau_v$ of the Earth’s core should be in the range between $2\times{10}^{35}$ $m^3 \cdot s$ to $6\times{10}^{36}\ m^3\cdot s$. By plotting these values and inspecting its intersection with hcp and bcc data in Fig. 4b, we obtain that the required undercooling is 470 K for bcc nucleation and 610 K for hcp nucleation. Therefore, the bcc phase significantly reduces the required undercooling of inner core nucleation by 140 K, which corresponds to ~1.4 billion years of cooling based on a cooling rate of 100 K/Gyr \cite{8}. 

\begin{figure}
\includegraphics[width=0.48\textwidth]{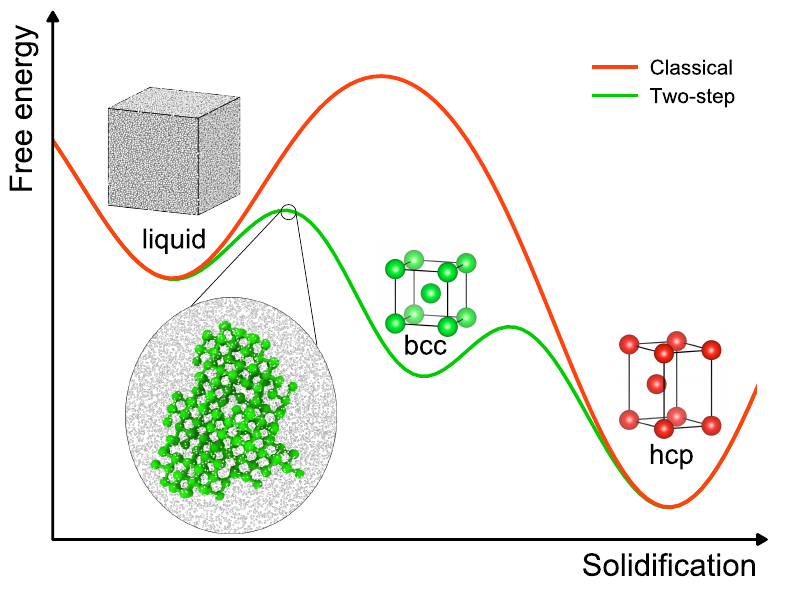}
\caption{\label{fig:fig5} \textbf{Schematics of the two-step nucleation process.}  The insert shows a bcc nucleus spontaneously formed during the brute-force MD simulations described in Supplementary Information Note S6.}
\end{figure}

\textbf{Discussion}
As illustrated in Fig. \ref{fig:fig5}, we have shown the bcc nucleation is likely to be the first step of iron crystallization under core conditions, which effectively decreases the large nucleation barrier of hcp phase. We note this two-step nucleation process can also be observed from the brute-force MD simulation by cooling the iron melts with ultra-high cooling rates (see Supplementary Information). While this cooling simulation is far from the core condition, it provides a qualitative validation of the PEM results. As the two-step nucleation process reduces the required undercooling of inner core nucleation, it can be a key to solving the core nucleation paradox.  While the necessary undercooling of the bcc phase (470 K) is still more significant than the inner core’s maxima undercooling of $\sim$200 K \cite{14}, a few factors remain to be accounted for. First, the core contains $\sim$10 wt.\% light elements. Davies \textit{et al.} \cite{16} have shown that oxygen can reduce the required undercooling by $\sim$55 K for the hcp nucleation. It was also found that the light elements can further stabilize the bcc phase w.r.t. the hcp phase \cite{4, 43, 44}. Therefore, their presence in the melt should cause a similar, if not larger, reduction of the required undercooling for the bcc nucleation. Moreover, the stresses associated with pressure variations can also reduce the undercooling by $\sim$100 K \cite{16}. Considering these effects, it should be possible to solve the nucleation paradox with future development of semi-empirical or machine-learning potentials for iron and light elements and more sophisticated PEM nucleation simulations of a multi-component melt. Our simulations consider homogenous nucleation only. While heterogeneous nucleation might decrease further the required undercooling, it is challenging to propose plausible scenarios to explain a preexisting stable substrate in the early deep core. This issue has been thoroughly discussed along with the nucleation paradox proposal in Huguet et al. \cite{14}. Besides, for any proposal of a preexisting substrate in the melt, the competition between bcc and hcp phases should still be considered to address the heterogeneous nucleation of the solid core. Then, our free energy results for bcc and hcp phases will also be relevant because the free energy barrier for heterogeneous nucleation, $\Delta G_{Het}$, includes the term $f\times \Delta G_{Hom}$, where $\Delta G_{Hom}$ is the homogenous nucleation barrier and f is the catalytic factor that depends on the contact angle, size, and substrate shape \cite{13}. Recent high-pressure and high-temperature experiments have reported the formation of bcc iron on the MgO surface \cite{45}, which indicates that the bcc should be the first nucleated phase in a heterogeneous nucleation process.

The two-step nucleation with the intermediate bcc phase can impact the present inner core structure. While the metastable bcc phase should eventually transform into the stable hcp phase at 323 GPa, the melting curves in Fig. \ref{fig:fig1}c suggest the bcc phase could be stabilized over the hcp phase with the increasing pressure when approaching the core center. The initially formed bcc phase in this region may remain at the core center nowadays. This coincides with hypotheses of a different innermost core structure proposed to explain the anomalous anisotropy of the inner core \cite{46,47,48}. Therefore, the two-step nucleation scenario of the inner-core formation opens a new path to understand the Earth’s deepest interior.

\section{Methods}
\textbf{\textit{Ab initio} molecular dynamics simulations} \textit{Ab initio} molecular dynamics (AIMD) simulations were performed to obtain the input data for the semi-empirical potential development. The Vienna \textit{Ab initio} Simulation Package (VASP) \cite{49} was employed for the density-functional theory (DFT) calculations. The projected augmented-wave (PAW) method was used to describe the electron-ion interaction, and the generalized gradient approximation (GGA) in the Perdew-Burke-Ernzerhof (PBE) form was employed for the exchange-correlation energy functional. The Mernin functional was used to equilibrate electrons and ions at the same temperature in AIMD simulations \cite{50}. This functional includes electronic entropic effects on the DFT energy.  The $\Gamma$  point was used to sample the Brillouin zone. The AIMD simulations were performed for the constant number of atoms, volume and temperature (NVT) ensemble. The Nos\'e-Hoover \cite{56} thermostat was employed to control the temperature. A time step of 2.0 $fs$ was used to integrate Newton’s equations of motion. Supercells with 288, 250, and 256 atoms were used to simulate hcp, bcc, and liquid models, respectively. To fit the potential for a large P-T range, three P-T conditions near the melting curves were investigated: 140 GPa at 4000 K, 250 GPa at 5500 K and 350 GPa at 6000 K. No phase transitions were observed during the simulation of either hcp, bcc or liquid model under these conditions during the AIMD simulations. To save time on the initial model equilibration, the AIMD simulations were run in an iterative manner with the potential development such that the available semi-empirical potential equilibrated the initial AIMD configurations. We monitored the energy and pressure as a function of time to determine when the models were equilibrated enough to start collecting data. Data collection took place in the last 10 $ps$ of each AIMD run. The potential was fitted to reproduce AIMD results using the Mermin functional, which includes electronic entropic effects.

\textbf{Classical Nucleation Theory formulas} Based on the CNT \cite{13}, the competition between the bulk and interface factors leads to the nucleation barrier. From Eq. (1), one can derive the nucleation barrier as
\begin{equation}
\Delta G^* = \frac{4s^3\gamma^3}{27 \Delta \mu^2 \rho_c^2} ,
\end{equation}
where $\rho_c$ is the crystal density and $s$ is a shape factor so that the interface area in Eq. (1) can be written as $A=s(\frac{N}{\rho_c})^{2/3}$. This equation can be simplified to Eq. (2) by replacing it with the critical nucleus size $N^\ast$
\begin{equation}
N^* = \frac{8s^3\gamma^3}{27 \lvert \Delta \mu \rvert ^3 \rho_c^3} ,
\end{equation}
From Eq. (4), one can derive the SLI free energy as
\begin{equation}
\gamma=\frac{3}{2s} \lvert \Delta \mu \rvert\rho_c^{\frac{2}{3}}N^{*\frac{1}{3}}. 
\end{equation}
The quantities to determine $\gamma$ can all be obtained from MD simulations.

\textbf{Persistent-embryo method}
The PEM \cite{28} is employed to measure the critical nucleus size $N^\ast$ with the classical MD simulations. It utilizes the main CNT concept that homogeneous nucleation happens via the formation of the critical nucleus in the undercooled liquid. During the simulation, a small crystal embryo containing $N_0$ atoms (should be much smaller than the critical nucleus) is constrained by spring forces to prevent melting \cite{28}. These forces are only applied to the original $N_0$ embryo atoms. The spring constant of the harmonic potential decreases with increasing nucleus size as $k(N)=k_0\frac{N_{sc}-N}{N_{sc}}$ if $N<N_{sc}$ and $k\left(N\right)=0$ otherwise. Here $N_{sc}$ is a subcritical threshold. No spring forces are applied if $N>N_{sc}$. This strategy ensures that the system is unbiased at the critical point such that a reliable critical nucleus can be obtained. If the nucleus melts below $N_{sc}$, the harmonic potential is gradually enforced, preventing the complete melting of the embryo. When the nucleus reaches the critical size, it has an equal chance to melt or to further grow, causing fluctuations around $N^\ast$. As a result, the $N(t)$ curve tends to display a plateau during the critical fluctuations, giving a unique signal to detect the appearance of the critical nucleus, as shown in Fig. 2a. The critical nucleus size is directly measured by averaging the nucleus size at the plateau \cite{28}. We repeated the PEM-MD simulation to collect at least four plateaus to obtain sufficient statistics and the confidence interval of the critical nucleus size $N^\ast$. The classical PEM-MD simulations were performed with the GPU (graphic processing unit)-accelerated LAMMPS (Large-scale Atomic/Molecular Massively Parallel Simulator) code \cite{51}. The interatomic interaction was modeled using the semi-empirical potential developed in this work based on the embedded atom method (EAM) \cite{52}. During the MD simulation, the constant number of atoms, pressure, and temperature (NPT) ensemble was applied with Nos\'e-Hoover thermostat and barostat. The damping time in the Nos\'e-Hoover thermostat was set as $\tau=0.1\ ps$ which is frequent enough for the heat dissipation during the crystallization \cite{42}. The time step of the simulation was 1.0 $fs$. In some cases where the nucleation was very fast, the time step was 0.8 $fs$. The simulation cell contained 31,250 atoms which are at least 15 times larger than the critical nucleus size. The simulation cell is sufficiently large to avoid the finite size effect (see Supplementary Information). The semi-empirical potential of iron developed in this work is publicly accessible at NIST potentials repository (https://www.ctcms.nist.gov/potentials/).

\section{Acknowledgments}

The authors are grateful to Prof. Dario Alfè and Prof. Anatoly Belonoshko for helping reproduce the previous semi-empirical potentials. This work was supported primarily by National Science Foundation awards EAR-1918134 and EAR-1918126. We acknowledge the computer resources from the Extreme Science and Engineering Discovery Environment (XSEDE), which is supported by National Science Foundation grant number ACI-1548562. R.M.W. and Y.S. also acknowledge partial support from the U.S. Department of Energy Grant DE-SC0019759. F.Z. acknowledges the support from U.S. Department of Energy, Basic Energy Sciences, Materials Science and Engineering Division, under Contract No. DEAC02-07CH11358.

\bibliographystyle{apsrev4-1}

\end{document}